\documentstyle[11pt]{article}
\begin{document}
\title{Gauge Field Model With Massive Gauge Bosons}
\author{{Ning Wu} \\
{\small Division 1, Institute of High Energy Physics, Box 918-1, 
Beijing 100039, P.R.China}}
\maketitle
\vskip 0.5in

\vskip 0.5in

\noindent
[Abstract] A gauge field model, which simultaneously has strict local gauge symmetry and massive 
gauge bosons, is discussed in this paper. The 
model has $SU(N)$ gauge symmetry. In the limit $\alpha \longrightarrow 0$, the gauge field model 
discussed in this paper will return to Yang-Mills gauge field model. The important meaning of this 
model is that, without Higgs mechanism, gauge bosons can also obtain masses. This theory can also 
explain some other unknown phenomenum. \\
\vskip 0.5in

~~~~ Since its foundation in 1954 \lbrack 1 \rbrack , the non-Abel  gauge field theory has been  
widely applied to elementary particle theory. Now, it is generally believed that the principle of local 
gauge invariance should play 
a fundamental role in the particles' interaction theory. We know that, if the gauge symmetry in 
the Yang-Mills theory is strictly preserved, the mass of the gauge field should be zero. But the  
mass of intermediate bosons which transmit weak interaction is very large. In order to construct a 
$SU(2)_L \times U(1)_Y$ unified electroweak gauge theory, we must use the Higgs mechanism 
\lbrack 2-4 \rbrack.  Up to now, almost every particle predicted by standard model was found by 
experiment except for Higgs particle. Does Higgs particle exist in nature? If there were no Higgs 
particle, how should the intermediate bosons $W^{\pm },Z$ obtain mass?\\

~~~~ The reason that we must introduce Higgs mechanism in the standard model is that, in the 
Yang-Mills theory, the mass of gauge bosons is zero. If we apply the Yang-Mills theory to the 
electroweak interactions, the mass of intermediate bosons is zero, which is contradict with 
experiment results. If the Yang-Mills theory is the only formalism of a gauge invariant theory, then, 
the introduction of the Higgs mechanism in the standard model can not be avoided and the 
introduction of Higgs particle can not be avoided as well. If, according to the principle of gauge 
invariance, we could construct some other gauge field models, then, the introduction of the Higgs 
mechanism in the standard model may be avoided and the introduction of Higgs particle may be 
avoided too. According to the principle of gauge invariance, we will construct a gauge field model in 
this paper, in which the local gauge symmetry is strictly preserved and the mass of gauge bosons is 
non-zero. In other words, 
without using Higgs particle, we could construct a gauge field model which have massive gauge 
bosons. This is the most important characteristics of the model and the most important difference 
between the Yang-Mills gauge field model and the model discussed in this paper. Applying the 
formalism given in this paper, we could construct fundamental particle theory to describe strong or 
electro-weak interactions \lbrack 5-6 \rbrack . When we use the gauge field model given in this 
paper 
to construct electroweak model, we will find that the introduction of the Higgs mechanism can be 
avoided and the introduction of Higgs particle can be avoided too. In other words, we could construct 
an electro-weak model which keeps most of the interaction properties of the standard model but has 
no Higgs particle. Besides, this new theory could explain some other unknown phenomena too 
\lbrack 
6 \rbrack. This means that Higgs particle may not exist in nature and physicists needn't spend so 
much effort and money in searching for it.
\\

~~~~For the sake of generality, let the gauge group be $SU(N)$ group. Suppose that $N$ fermion 
fields $\psi_l (x)~~(l=1,2, \cdots N)$ form a multiplet of matter fields. Denote:\\
$$
\psi (x) =\left ( 
\begin{array}{c}
\psi_1 (x)\\
\psi_2 (x)\\
\vdots\\
\psi_N (x)
\end{array}
\right ) \eqno{(1)} 
$$
All $\psi (x)$ form a representative space of $SU(N)$ group. In this space, we denote the 
representative matrices of the generators of $SU(N)$ group by $T_i ~~(i=1,2, \cdots N^2-1) $. They 
satisfy
$$
\lbrack T_i ~ ,~ T_j \rbrack = i f_{ijk} T_k
\eqno{(2)} 
$$
$$
Tr( T_i  T_j ) = \delta_{ij} K.
\eqno{(3)} 
$$
The representative matrix of a general element of the $SU(N)$ group can be written as:
$$
U = e^{-i \alpha ^i T_i }
\eqno{(4)}
$$
with $\alpha ^i$ the real group parameters. $U$ is a unitary $N \times N$ gauge transformation 
matrix
$$
U^{\dag}U=1=U U^{\dag}.
\eqno{(5)} 
$$

~~~~We will introduce two sets of gauge fields denoted by $A_{\mu} (x)$ and $B_{\mu} (x)$. they 
can be expressed as:
$$
A_{\mu}(x) = A_{\mu} ^i (x) T_i
\eqno{(6)} 
$$
$$
B_{\mu}(x) = B_{\mu} ^i (x) T_i,
\eqno{(7)}
$$
where summation convention of repeated index is used. Correspondingly, we introduce two gauge 
covariant derivatives
$$
D_{\mu} = \partial _{\mu} - ig A_{\mu}
\eqno{(8)} 
$$
$$
D_{b \mu} = \partial _{\mu} + i \alpha g B_{\mu}
\eqno{(9)} 
$$
The strengthes of two gauge fields are respectively defined as:
$$
A_{\mu \nu} = \frac{1}{-ig} \lbrack D_{\mu} ~,~ D_{\nu} \rbrack = A_{\mu \nu}^i T_i
\eqno{(10)} 
$$
$$
B_{\mu \nu} = \frac{1}{i \alpha g} \lbrack D_{b\mu} ~,~ D_{b\nu} \rbrack = B_{\mu \nu}^i T_i
\eqno{(11)} 
$$
where 
$$
A_{\mu \nu}^i = \partial _{\mu} A_{\nu}^i - \partial _{\nu} A_{\mu}^i
+g f^{ijk} A_{\mu}^j    A_{\nu}^k
\eqno{(12)} 
$$
$$
B_{\mu \nu}^i = \partial _{\mu} B_{\nu}^i - \partial _{\nu} B_{\mu}^i
- \alpha g f^{ijk} B_{\mu}^j    B_{\nu}^k
\eqno{(13)} 
$$

~~~~The lagrangian of the model is :
$$
\begin{array}{ccl}
\cal L &= &- \overline{\psi}(\gamma ^{\mu} D_{\mu} +m) \psi 
-\frac{1}{4K} Tr( A^{\mu \nu} A_{\mu \nu} )
-\frac{1}{4K} Tr( B^{\mu \nu} B_{\mu \nu} ) \\
&&-\frac{\mu ^2}{2K ( 1+ \alpha ^2)} 
Tr \left \lbrack (A^{\mu}+\alpha B^{\mu})( A_{\mu}+\alpha B_{\mu} ) \right \rbrack
\end{array}
\eqno{(14)} 
$$
with $\alpha$ a non-negative real constant. In this paper, the apace time metric is selected as 
$\eta _{\mu \nu}=diag(-1,1,1,1) ~~ (\mu,\nu =0,1,2,3)$

~~~~It is easy to prove that the above lagrangian is invariant under the following local gauge 
transformations:
$$
\psi \longrightarrow U \psi
\eqno{(15)} 
$$
$$
A_{\mu} \longrightarrow U A_{\mu} U^{\dag}
-\frac{1}{ig}U \partial _{\mu}U^{\dag}
\eqno{(16)} 
$$
$$
B_{\mu} \longrightarrow U B_{\mu} U^{\dag}
+\frac{1}{i \alpha g}U \partial _{\mu}U^{\dag}.
\eqno{(17)} 
$$
So, The model has local gauge symmetry.\\

~~~~The gauge field $A_{\mu}$ and  $B_{\mu}$ are not eigenvectors of mass matrix. In order to 
construct the eigenvectors of mass matrix, let's define the following transformations:\\
$$
C_{\mu}={\rm cos}\theta A_{\mu}+{\rm sin}\theta B_{\mu}
\eqno{(18)} 
$$
$$
F_{\mu}=-{\rm sin}\theta A_{\mu}+{\rm cos}\theta B_{\mu},
\eqno{(19)} 
$$
where
$$
{\rm cos}\theta = \frac{1}{\sqrt{1+\alpha ^2}} ~,~
{\rm sin}\theta = \frac{\alpha}{\sqrt{1+\alpha ^2}} 
\eqno{(20)} 
$$
Then, the lagrangian $\cal L$ given by eq.(14) changes into:
$$
{\cal L}={\cal L}_0 + {\cal L}_I,
\eqno{(21)} 
$$
$$
{\cal L}_0= - \overline{\psi}(\gamma ^{\mu} \partial _{\mu} +m) \psi 
-\frac{1}{4} C^{i \mu \nu}_0 C^i_{0 \mu \nu} 
-\frac{1}{4} F^{i \mu \nu}_0 F^i_{0 \mu \nu}
-\frac{\mu ^2}{2} C^{i \mu} C^i_{\mu}.
\eqno{(22)} 
$$
${\cal L}_I$ is interaction lagrangian which only contains interaction terms. In eq.(22), we have 
used 
the following notes:
$$
C_{0 \mu \nu}^i = \partial _{\mu} C_{\nu}^i - \partial _{\nu} C_{\mu}^i
\eqno{(23)} 
$$
$$
F_{0 \mu \nu}^i = \partial _{\mu} F_{\nu}^i - \partial _{\nu} F_{\mu}^i
\eqno{(24)} 
$$
From eq.(22), we know that the mass of gauge field $C_{\mu}$ is $\mu$, and the mass of gauge 
field $F_{\mu}$ is zero. So, there exist a massive gauge field and a massless gauge field 
simultaneously in the model.\\

~~~~If $\alpha$ is small enough, that is
$$
\alpha \ll 1,
\eqno{(25)} 
$$
then, in the leading term,
$$
{\rm cos}\alpha \approx 1 ~,~  {\rm sin}\alpha \approx 0,
\eqno{(26)} 
$$
$$
A_{\mu} \approx C_{\mu} ~,~  B_{\mu} \approx F_{\mu}.
\eqno{(27)} 
$$
In this case, the lagrangian $\cal L$ changes into:
$$
\begin{array}{ccl}
\cal L & \approx &- \overline{\psi} \lbrack \gamma ^{\mu} (\partial{\mu}-i g C_{\mu} )+m \rbrack 
\psi 
-\frac{1}{4} F^{i \mu \nu}_0  F_{0 \mu \nu}^i \\
&&-\frac{1}{4} C^{i \mu \nu} C^i_{\mu \nu} -\frac{\mu ^2}{2} C^{i \mu} C^i_{\mu} ~,
\end{array}
\eqno{(28)} 
$$
where
$$
C_{\mu \nu}^i = \partial _{\mu} C_{\nu}^i - \partial _{\nu} C_{\mu}^i
+g f^{ijk} C_{\mu}^j    C_{\nu}^k.
\eqno{(29)} 
$$
Now, we see that, except for a gauge field mass term and a dynamically term of $F^i_{\mu}$ which 
doesn't interact with matter field $\psi$ in the leading term, this lagrangian is the same as that of 
Yang-Mills gauge theory. So, we could anticipate that these two gauge theories will give similar 
dynamical behavior in describing particles'  interaction.\\

~~~~If we apply this model to QCD \lbrack 5 \rbrack, we will obtain two sets of gluons, one set is 
massive gluons, another is massless gluons. Because $SU(3)_c$ symmetry is a strict symmetry,  
these gluons are confinement. The important thing is that there may exists three sets of glueballs 
which may have different masses but have the same spin-parity. If we apply this model to electro-
weak interactions \lbrack 6 \rbrack , we will get two sets of gauge bosons that transmit electro-weak 
interactions, one set is massive that has been found by experiment, another is massless that is not 
found by experiment. The existence of these massless gauge bosons could be used to explain some 
unknown phenomenon \lbrack 5-6 \rbrack. If $\alpha$ is 
very small, the coupling between these massless gauge bosons and leptons or quarks will be very 
weak. Therefore the model discussed in this paper does not contradict with the present experiment.\\

~~~~Finally, we should say that, without Higgs mechanism, we could also construct a electro-weak 
model in which the mass of intermediate gauge bosons is non-zero. And at a appropriate limit, the 
model will go back to the standard model \lbrack 6 \rbrack. \\

\section*{Reference:}
\begin{description}
\item[\lbrack 1 \rbrack]  C.N.Yang, R.L.Mills, Phys Rev {\bf 96} (1954) 191
\item[\lbrack 2 \rbrack]  S.Glashow, Nucl Phys {\bf 22}(1961) 579
\item[\lbrack 3 \rbrack]  S.Weinberg, Phys Rev Lett {\bf 19} (1967) 1264
\item[\lbrack 4 \rbrack]  A.Salam, in Elementary Particle Theory, eds.N.Svartholm(Almquist and 
Forlag, Stockholm,1968)
\item[\lbrack 5 \rbrack]  Ning Wu, A new QCD model  (in preparation)
\item[\lbrack 6 \rbrack]  Ning Wu, A new unified electro-weak model  (in preparation)
\end{description}

\end{document}